\documentclass[journal]{IEEEtran}
\IEEEoverridecommandlockouts
\usepackage{cite}
\usepackage{amsmath,amssymb,amsfonts,amsthm}

\usepackage[utf8]{inputenc} 
\usepackage[T1]{fontenc}

\usepackage[normalem]{ulem}

\usepackage[inline,shortlabels]{enumitem}
\setlist[enumerate]{nosep}
\setlist{itemjoin ={,\enspace}, itemjoin* = {, and\enspace}}

\usepackage{graphicx}
\usepackage{textcomp}
\usepackage{xcolor,soul}
\usepackage{stmaryrd}
\usepackage{algorithm}
\usepackage{algpseudocode}
\usepackage{amsmath}
\usepackage{mathrsfs}
\usepackage{bm}
\usepackage[abs]{overpic}
\usepackage{dsfont}

\usepackage{physics}
\usepackage{tikz}
\usepackage{mathdots}
\usepackage{yhmath}
\usepackage{cancel}
\usepackage{siunitx}
\usepackage{array}
\usepackage{multirow}
\usepackage{amssymb}
\usepackage{gensymb}
\usepackage{tabularx}
\usepackage{extarrows}
\usepackage{booktabs}
\usetikzlibrary{fadings}
\usetikzlibrary{patterns}
\usetikzlibrary{shadows.blur}
\usetikzlibrary{shapes}

\theoremstyle{plain}
\newtheorem*{theorem*}{Theorem}
\newtheorem{theorem}{Theorem}

\newtheorem*{eg}{Example}

\usepackage{xcolor,varwidth}

\usepackage[scaled=.75]{beramono}
\newcommand{\bpara}[1]		{\medskip \noindent {\bf #1}}

\renewcommand\geq\geqslant
\renewcommand\leq\leqslant

\def\DE					{\stackrel{\rm{def}}{=}}
\newcommand\rob[1]			{\left( #1 \right)}

\newcommand\sqb[1]			{\left[ #1 \right]}
\newcommand\ym[2]			{#1_{\lambda}\sqb{#2}}

\newcommand\fig[1]			{Fig.~\ref{#1}}

\def\qo                     {q_0}

\def\eg					    {\emph{e.g.,}~}
\def\ie					    {\emph{i.e.,}~}

\usepackage{xspace}

\def\madc					{{$\mathscr{M}_\lambda$-{\fontsize{11pt}{11pt}\selectfont\texttt{ADC}}}\xspace}

\def\madcs					{{$\mathscr{M}_\lambda$-{\fontsize{11pt}{11pt}\selectfont\texttt{ADCs}}}\xspace}

\DeclareDocumentCommand{\sd}{o}  
{{\underline{\ast}\IfValueT{#1}{_{#1}}}}

\DeclareDocumentCommand{\xmod}{m o o o}  
{%
\IfNoValueTF{#4}
{{#1}\IfValueT{#2}{_{m,\mathsf{#2}}}\IfValueT{#3}{#3}}
{{#1}\IfValueT{#2}{_{m,\mathsf{#2}}^{\mathsf{#4}}}\IfValueT{#3}{#3}}
}

\newcommand{\MO}[1]		{\mathscr{M}_\lambda ({#1} )}
\newcommand{\VO}[1]		{\varepsilon_{#1}}

\newcommand{\BL}[1] 		{\mathcal{B}_{\Omega}}

\newcommand{\ft}[1]			{\left[\kern-0.15em\left[#1\right]\kern-0.15em\right]}
\newcommand{\fe}[1]		{\left[\kern-0.30em\left[#1\right]\kern-0.30em\right]}

\DeclareMathAlphabet{\mathsfit}{T1}{\sfdefault}{\mddefault}{\sldefault}
\SetMathAlphabet{\mathsfit}{bold}{T1}{\sfdefault}{\bfdefault}{\sldefault}
\usepackage[switch]{lineno}
\usepackage{siunitx}

\def\BibTeX{{\mathrm B\kern-.05em{\sc i\kern-.025em b}\kern-.08em
    T\kern-.1667em\lower.7ex\hbox{E}\kern-.125emX}}

\begin{document}
\title{Full-Duplex Beyond Self-Interference: \\ The Unlimited Sensing Way
}

\author{Ziang Liu,
        Ayush Bhandari,
        and~Bruno Clerckx,~\IEEEmembership{Fellow,~IEEE}        
\thanks{The authors are with the Dept. of Electrical and Electronic Engg., Imperial College London, SW7 2AZ, UK. (e-mails:\{ziang.liu20, a.bhandari, b.clerckx\}@imperial.ac.uk).} }


\maketitle

\begin{abstract}
The success of full-stack full-duplex communication systems depends on how effectively one can achieve digital self-interference cancellation (SIC). Towards this end, in this paper, we consider unlimited sensing framework (USF) enabled full-duplex system. We show that by injecting folding non-linearities in the sensing pipeline, one can not only suppress self-interference but also recover the signal of interest (SoI). This approach leads to novel design of the receiver architecture that is complemented by a modulo-domain channel estimation method. We then demonstrate the advantages of \madc by analyzing the relationship between quantization noise, quantization bits, and dynamic range. Numerical experiments show that the USF enabled receiver structure can achieve up to $40$ dB digital SIC by using as few as $4$-bits per sample. Our method outperforms the previous approach based on adaptive filters when it comes to digital SIC performance, SoI reconstruction, and detection.

\end{abstract}


\section{Introduction}
Future growth of wireless communication systems in terms of applications and transmission bandwidth poses strong restrictions on the frequency spectrum \cite{yang20196g}. 
A promising solution to the \emph{spectrum scarcity}  problem has been the design of In-Band Full-Duplex (FD) systems \cite{sabharwal2014band}. It has attracted a surge of research interest due to potential doubling of the spectral efficiency \cite{sabharwal2014band}. However, the coupling between the co-located transmitter and receiver in FD systems causes severe self-interference (SI) (\eg up to $100$ dB \cite{sabharwal2014band}). This results in an undesirable scenario where the signal of interest (SoI) containing the uplink data irretrievably drowns in the interference. 
On the one hand, simultaneous occurrence of SI and SoI naturally necessitates a high dynamic range (HDR) analog-to-digital converters or ADCs, or else, there will be no hope of recovering the uplink data from this occurrence. On the other hand, for a fixed bit budget, HDR ADCs entail higher quantization noise which can compromise the capability of existing algorithms. In this case, the only available remedy is to boost the number of bits used for digitization. This strategy, proves to be sub-optimal because higher bit-budget entails higher power consumption and cost. 
In crux, combating the effect of SI creates an inevitable trade-off between the bit-budget and the dynamic range (DR).





%
To mitigate SI, a straightforward approach is to subtract the reconstructed SI from the received signal in the digital domain, given that the transmitted signal is known. However, the primary challenge lies in the DR of the ADC, which is typically smaller than that of the received signal. As a result, clipping errors and significant quantization noise occur, leading to irreversible information loss. Therefore, SIC in the analog domain is necessary to ensure that the received signal's DR fits within the ADC's DR \cite{sabharwal2014band}. For instance, a two-stage analog SIC architecture achieving a total of 79 dB suppression is proposed in \cite{quan2017two}. Unlike conventional SIC methods, we aim to directly cancel the SI in the digital domain by addressing the limitations imposed by the ADC's DR. This approach simplifies the SIC pipeline by reducing hardware complexity.

This new approach is realized through the Unlimited Sensing Framework (USF), which changes the FD system pipeline by jointly designing hardware and algorithms. This integration breaks the  trade-off between DR and digital resolution \cite{Bhandari:2020:Ja}. Specifically, the analog-domain modulo folding, implemented via a modulo ADC, prevents ADC saturation, allowing HDR signals to be preserved with minimal quantization noise. Advanced signal processing algorithms are then employed to decode the folded signal, offering unconventional advantages. Recently, the USF has gained significant attention in both theory and applications. In \cite{guo2023iter}, a robust algorithm was proposed based on optimization. A receiver design for massive MIMO systems based on the USF was proposed in \cite{liu2023lambda} to prevent ADC saturation (\eg in the near-far user problem) and to increase communication rates. Unlike \cite{liu2023lambda}, the design in this paper addresses the FD case, introducing new aspects such as modulo-domain sparse channel estimation and digital SIC. We validate our approach through simulations, demonstrating its superiority over recent methods. Our approach achieves 40 dB digital SIC with as few as 4-bit quantization resolution.
 \bpara{Contributions.} Our main contributions are as follows:  
 \begin{enumerate}[leftmargin = *,label = \tiny$\bullet$]
    \item \textbf{Theoretical front.}  We propose a \emph{full-stack} FD receiver architecture design, which can handle HDR signals and achieve higher quantization resolution with a given budget. The \emph{full-stack} design in this paper enables understanding of the trade-off between bit-budget and DR, and the best achievable digital SI cancellation capability.

    \item \textbf{Algorithmic front.} We propose a novel  modulo-domain sparse channel estimation method, resulting in a more accurate digital SI cancellation compared with previous methods based on adaptive filters \cite{ordonez2021full}.

    \item \textbf{Hardware front.}
    Since our approach is devoid of any analog-domain SIC circuitry, \eg AGC and VGA, it lends itself to a low complexity \textit{hardware} implementation.

 \end{enumerate}
Ordonez \& co-workers were early to identify the  benefits of modulo non-linearties for combating receiver saturation in a FD setup \cite{ordonez2021full}. Their work presents a systematic analysis of ``Modulo FD Transceiver'' inspiring new architectures in the field. Compared to \cite{ordonez2021full}, our work has fundamental differences:
 \begin{enumerate}[leftmargin = *, label = \arabic*.]
    \item Our scope includes the \textit{recovery problem} in the \textit{full-stack} FD architecture, unlike \cite{ordonez2021full} which stops at the ADCs and does not cover subsequent processes. This is relevant and timely for realizing a fully digital-domain FD design.

    \item We propose a new modulo-domain channel estimation method instead of the adaptive filter-based method in \cite{ordonez2021full}. Additionally, we provide a sampling theorem that ensures the success of our approach, enabling accurate SI channel estimation and fully digital-domain SI cancellation.
    

    \item We consider low-resolution quantization to facilitate lower power consumption, as ADC quantization dominates power usage. In contrast, \cite{ordonez2021full} does not account for this advantage.

\end{enumerate}

\begin{figure}[t]
\centering
		\includegraphics[width = 0.45\textwidth]{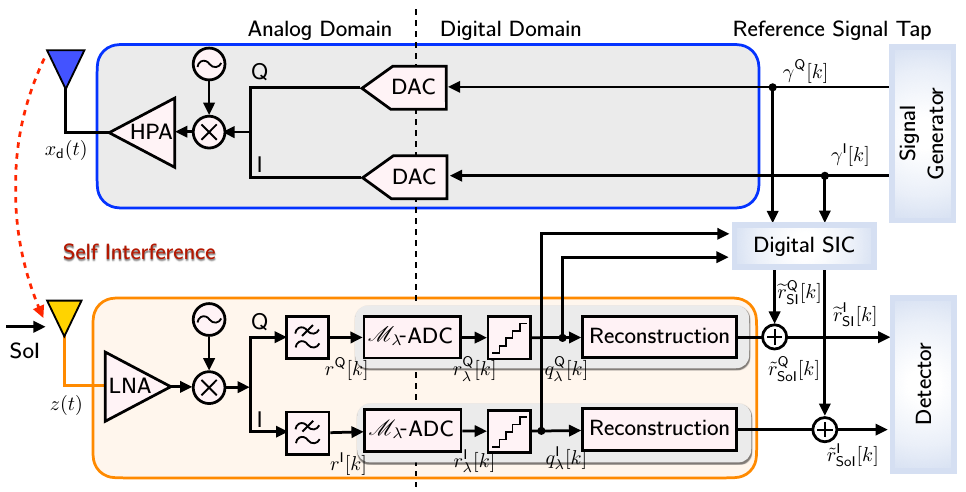}
		\caption{FD model with modulo-domain channel estimation method and digital cancellation based on \madc.}
		\label{fig:system_model}
\end{figure}


\section{System model}
The FD communication system we consider is shown in \fig{fig:system_model}. The baseband signals, $\gamma^\mathsf{Q}[k]$ and $\gamma^\mathsf{I}[k]$, are pulse-shaped and up-converted to form the passband signal, $x_\mathsf{d}(t)$. The FD transceiver simultaneously serves an uplink and a downlink user. The received antenna signal is given by
\vspace{-3pt}
\begin{equation}
\label{eq:ztime}
z(t) \! \DE \! \underbrace{{ \sqrt{p_\mathsf{u}} \left( {{h_\mathsf{u}}{\,{\underline  *  }_T\,}{x_\mathsf{u}}} \right)} \left( t \right)}_\textsf{SoI} \, +  \, \underbrace{{ \sqrt{p_\mathsf{d}} \left( {{h_\mathsf{SI}}{\,{\underline  *  }_T\,}{x_\mathsf{d}}} \right)} \left( t \right)}_\textsf{SI}  \, +  \, n(t),
\vspace{-3pt}
\end{equation}
\begin{enumerate}[leftmargin = *, label = --$\bullet$]
    \item $\sd[T]$ is the semi-discrete convolution operator with sampling interval $T$ and some sequence $c \in \mathbb{\ell}^2$, defined by
    \vspace{-3pt}
    \[
    \sd[T]: c \longmapsto \rob{c \sd[T] \phi}\rob{t} =\sum\nolimits_{n\in\mathbb{Z}}c\sqb{n} \phi\rob{t-kT}.
    \vspace{-3pt}
    \]

    \item $p_\mathsf{u}$ and $p_\mathsf{d}$ are the transmit power of the user and BS, respectively. $h_\mathsf{u}$ and $h_\mathsf{SI}$ are the discrete uplink and SI channels in the FD system, respectively. 
    \item $x_\mathsf{u}(t)$ and $x_\mathsf{d}(t)$ are the pulse-shaped continuous-time uplink signal and downlink signal, respectively.
    \item $n(t) \sim \mathcal{C N}(0, 1)$ is the additive complex-Gaussian noise at the receiver.
\end{enumerate}
Subsequently, the received signal enters the RF chain and is processed by the \madc at the quadrature and in-phase channels, respectively. As shown in \fig{fig:system_model}, we design a novel RF chain, signal processing blocks for SoI reconstruction, digital SIC, and detection. 

\bpara{Receiver.}
Compared with conventional receiver architecture for FD system, our novel receiver deploys a \madc to simultaneously prevent drowning of SoI caused by the saturation due to the limited DR, and reduce quantization noise. Specifically, the continuous-time baseband signal for $\mathsf{I/Q}$ channels, $r^{\mathsf{Q}}(t)$ and $r^{\mathsf{I}}(t)$, are first folded with modulo threshold $\lambda>0$ by the \madc hardware \cite{Bhandari:2021:J}. 
Since, 
\begin{enumerate*}[label = \uline{\roman*}) ]
    \item the modulo signal $r(t)$ is bounded by $\lambda$ (in absolute sense), no VGA and AGC \cite{sheemar2020receiver} is needed
    \item there is no requirement for analog-domain SIC circuitry,
\end{enumerate*}
our approach results in a lower hardware complexity. The continuous-time modulo-folded signal is sampled and quantized, resulting in, 
\begin{equation}
r(t) \xrightarrow{{{\text{Folding}}}}\boxed{{\mathscr{M}_\lambda}} \to \ym{r}{k} 
\xrightarrow{{{\text{b Bits}}}}\boxed{{\mathscr{Q}_b}} \to 
\ym{q}{k} = \mathscr{Q}_b(\ym{r}{k})
\label{eq:mody}
\nonumber
\end{equation}

Formally, the centered modulo operator \cite{Bhandari:2020:J,liu2023lambda} is defined by,
\vspace{-3pt}
\begin{equation}
\mathscr{M}_{\lambda}: r \mapsto 2 \lambda\left(\biggr \llbracket \frac{r}{2 \lambda}+\frac{1}{2} \biggr \rrbracket-\frac{1}{2}\right),
\vspace{-3pt}
\label{eq:modulo}
\end{equation}
where $\llbracket r \rrbracket \stackrel{\text { def }}{=} r-\lfloor r\rfloor$ is the fractional part of  $r$.

Additionally, $\ym{r}{k} \stackrel{\text { def }}{=} {\left. \mathscr{M}_\lambda(r(t)) \right|_{t = kT}}$ and $\mathscr{Q}_b$ is the quantization operator with bit budget of $b$ bits. These quantized samples $\ym{q}{k}$ enable the reconstruction of the SI $r_\mathsf{SI}$, and thus we can achieve SIC by subtracting it from the received signal $r$. {We assume a perfect synchronization by a cross-correlation method \cite{li2018self}.}

\bpara{Recovery from folded samples.}
To reconstruct the HDR input signal, $r(t)$, from its modulo samples, an "inversion" of modulo non-linearity $\mathscr{M}_\lambda{\rob{\cdot}}$ is needed. The key to unfolding the modulo samples is based on the observation that the finite difference operator commutes with the modulo in some sense \cite{Bhandari:2020:Ja}. This allows us to obtain the higher-order differences of $r$ from its modulo samples, $r_\lambda$. Thus, the reconstruction amounts to the problem of stably inverting the difference-operator.

Let us define the $L^\textrm{th}$ order finite difference of the sequence $x[k]$ by the recursion, given by
\vspace{-3pt}
\begin{equation}
    \left(\Delta^{L} x\right)[k]=\Delta^{L-1}(\Delta x)[k], L > 1.
\vspace{-3pt}
\end{equation}
When $L = 1$, we define 
\vspace{-3pt}
\begin{equation}
    \underline{x}[k] = (\Delta x)[k] {=} x[k+1]-x[k]  \iff \underline{\mathbf{x}} = \boldsymbol{\Delta} \mathbf{x}.
\vspace{-3pt}
\end{equation}
Specifically, according to the unlimited sampling theorem with bounded noise \cite{Bhandari:2020:Ja},
oversampling $r\rob{t}\in \BL{}$ with $\mathrm{T} \leqslant {1}/{2^\alpha \Omega e}$, where $\alpha \in \mathbb{N}$, $\Omega$ is the bandwidth, and $e$ is the Euler's number, allows us to access the $L^\text{th}$ order difference, $\Delta^L r$, from its modulo samples $r_\lambda\sqb{k}$. A suitable $L$ can be 
\vspace{-3pt}
\begin{equation}
    {L^ \star } \geqslant \left\lceil {\frac{{\log \lambda  - {\beta _r}}}{{\log \left( {T\Omega e} \right)}}} \right\rceil \text{with} \, \beta _r \in 2 \lambda \mathbb{Z} \,\, \text{and} \,\, \beta _r \geq \|r\|_\infty
\vspace{-3pt}
\end{equation}

To stably invert the difference operator, we use the anti-difference operation $\mathsf{S}$, defined by
\vspace{-10pt}
\begin{equation}
\mathsf{S}:\{s[k]\}_{k \in \mathbb{Z}^{+}} \rightarrow \sum_{m=1}^k s[m].
\vspace{-3pt}
\end{equation}
Instead of recovering $r$ via anti-difference operation $\mathsf{S}$, we capitalize the piece-wise constant structure of the residue $\VO{r}\rob{t} = r\rob{t} - \MO{r\rob{t}}$, that is, $\Delta^L\VO{r}\rob{kT}\in2\lambda\mathbb{Z}$, to reconstruct the unfolded signal.
The reconstruction approach is the same as in \cite{liu2023lambda}.

\bpara{Modulo-domain channel estimation and digital SIC.} To achieve a full-stack fully digital-domain SIC FD system, we propose a modulo-domain channel estimation method to estimate SI channel, $h_\mathsf{SI}$, as in Algorithm \ref{alg:FPrecovery}. In addition, we design a corresponding digital SIC approach based on \madc, inspired by \cite{Bhandari:2020:Ja, bhandari2022back}. Since the received signal can be acquired without saturation, and it incorporates SI and SoI as given in \eqref{eq:ztime}, our aim is to reconstruct SI via our proposed approach and subtract it from the received signal.

Specifically, in the proposed modulo-domain channel estimation method, a pilot is firstly transmitted, and we assume no SoI is transmitted during the SI channel estimation period. This is the typical configuration for SI channel estimation \cite{korpi2015achievable}. Considering a single path SI channel, $h_\mathsf{SI}$, with delay, $\tau$, and amplitude, $A$, the SI channel is hence modelled as $h_\mathsf{SI}[k] = A \delta[k-\tau] $\footnote{We assume a single-path self-interference (SI) channel in the initial stage of our research. Future work will include the exploration of more realistic scenarios, such as multipath and time-varying channel models.}.
After down-conversion, the pointwise received SI samples are given by
\vspace{-3pt}
\begin{equation}
\label{eq:r_time}
    r_\mathsf{SI}[k] = A \gamma[k-\tau]
    = \sum\nolimits_{|p| \leq P} \left(\widehat{\gamma}_p \widehat{h}_\mathsf{SI}\right) e^{\jmath p \omega_0 k}, 
    \vspace{-3pt}
\end{equation}
where $\gamma$ is a $T_p$-periodic baseband transmit signal (reference signal, cf. transmitter in \fig{fig:system_model}), $\widehat{h}_\mathsf{SI}= A e^{-\jmath p \omega_0 \tau}$, $P = \lceil \Omega/\omega_0 \rceil$, $\omega_0 = 2 \pi /T_p$, and $K \geq 2P+1$. Additionally, $\widehat{\gamma}_p$ is the Fourier series (FS) coefficient and $\widehat{\gamma}_{-p} = \widehat{\gamma}^*_{p}$. We then have the bandlimitedness that is $\widehat{\gamma}_p = 0, |p| > P = [\Omega/\omega_0]$ denoted by $f \in \mathcal{B}_\Omega$. In vector-matrix format, \eqref{eq:r_time} is given by
\begin{equation}
\label{eq:rsimat}
    \mathbf{r}_\mathsf{SI} = \mathbf{U}^* \,\boldsymbol{\mathcal{D}}_{\widehat{\gamma}} \, \widehat{\mathbf{h}}_\mathsf{SI},
    \vspace{-3pt}
\end{equation}
\begin{enumerate}[leftmargin = *, label =--$\bullet$]
    \item $\mathbf{r}_\mathsf{SI} \in \mathbb{R}^{K \times 1}$ is the samples with $[\mathbf{r}_\mathsf{SI}]_k = r_\mathsf{SI}({k}T)$.
    \item $\mathbf{U} \in \mathbb{C}^{K \times (2P+1)}$ is the Discrete Fourier Transform (DFT) matrix with element $[\mathbf{U}]_{k,p} = e^{- \jmath p \omega_0 k T}$.
    \item $\boldsymbol{\mathcal{D}}_{\widehat{\gamma}}$ is the matrix of FS coefficients. It is a diagonal matrix and the FS coefficients of the kernel (transmit signal), $\gamma$, are on the diagonal, \ie $[\boldsymbol{\mathcal{D}}_{\widehat{\gamma}}]_{p,p} = \widehat{\gamma}_p$.
    \item $\widehat{\mathbf{h}}_\mathsf{SI} \in \mathbb{C}^{(2P+1) \times 1}$ is a vector of exponentials parameterized by the unknown channel, \ie $\widehat{\mathbf{h}}_{\mathsf{SI}, p} = \widehat{h}_{\mathsf{SI}, p}$.
\end{enumerate}
In our case, the forward model takes the form of,
\begin{equation}
     \mathbf{r}_\mathsf{SI, \lambda} =  \mathscr{M}_\lambda (\mathbf{U}^* \,\boldsymbol{\mathcal{D}}_{\widehat{\gamma}} \, \widehat{\mathbf{h}}_\mathsf{SI}) \in \mathbb{R}^{K \times 1},
\end{equation}
and the goal is to estimate $\mathbf{r}_\mathsf{SI}$ in \eqref{eq:rsimat}.

\bpara{Estimation Procedure.} Starting from the modulo decomposition property, we have 
\vspace{-3pt}
\begin{equation}
\label{eq:rsi_lambda}
    \mathbf{r}_\mathsf{SI, \lambda} = \mathbf{r}_\mathsf{SI} - \boldsymbol{\varepsilon}_\mathsf{SI}
    \xrightarrow{\Delta}        \underline{\mathbf{r}}_\mathsf{SI, \lambda} = \underline{\mathbf{r}}_\mathsf{SI} - \underline{\boldsymbol{\varepsilon}}_\mathsf{SI}, \quad 
    \underline{\mathbf{r}}_\mathsf{SI} = \Delta\mathbf{r}_\mathsf{SI} 
\vspace{-3pt}
\end{equation}
where $\Delta$ is the finite difference operator yielding, 
$[\Delta\mathbf{r}_\mathsf{SI}]_k =  [\mathbf{r}_\mathsf{SI}]_{k+1} - [\mathbf{r}_\mathsf{SI}]_k$.
We use $\mathbb{I}_{K} = \{0, \cdots, K-1 \}, K \in \mathbb{Z}^+$ to denote the set of $K$ contiguous integers, and $\mathbb{E}_{P, K}=[0, P] \cup[K-P, K-1],\left|\mathbb{E}_{P, K}\right|=2 P+1$ to denote the effective spectrum.
Applying the DFT on \eqref{eq:rsi_lambda} results in,
\vspace{-3pt}
\begin{equation}
\label{eq:rsi_lambda_finite_dft}
 \mathbf{V} \underline{\mathbf{r}}_{\mathsf{SI}, \lambda} \! = \! \mathbf{V} \underline{\mathbf{r}}_\mathsf{SI} \! - \! \mathbf{V} \underline{\boldsymbol{\varepsilon}}_\mathsf{SI}, \, \, [\mathbf{V}]_{n, k}=e^{-\jmath\left(\frac{2 \pi}{K-1}\right) n k}, k \in \mathbb{I}_{K-1}
 \vspace{-3pt}
\end{equation}
where $\mathbf{V} \in \mathbb{C}^{(K-1) \times (K-1)}$ is the DFT matrix. Noting that $\mathbf{V} \underline{\mathbf{r}}_\mathsf{SI}$ is the contribution due to the bandlimited part, choosing 
$n \in\left(\mathbb{I}_{K-1} \backslash \mathbb{E}_{P, K-1}\right)+(K-1) \mathbb{Z}$ provides us with access to the modulo folds, which in the DFT domain assume a parametric form. In particular, $\forall n \in\left(\mathbb{I}_{K-1} \backslash \mathbb{E}_{P, K-1}\right)+(K-1) \mathbb{Z}$,  $\mathbf{V} \underline{\mathbf{r}}_{\mathsf{SI}, \lambda} = -\mathbf{V} \underline{\varepsilon}_\mathsf{SI} = \sum_{m=0}^{M-1} \mu_m e^{-\jmath\left(\frac{2 \pi}{K-1}\right) \frac{n}{T} \nu_m}$ is sum-of-complex-exponentials \cite{bhandari2022back, Bhandari:2021:J} which can be estimated via  Prony's method \cite{stoica1993list}. We denote this procedure by, \texttt{Prony}$(\mathbf{z}) \mapsto\left(\left\{\mu_m, \nu_m\right\}_{m=0}^{M-1}, \mathbf{f}\right)$ where $\mathbf{z} = -\mathbf{V} \underline{\varepsilon}_\mathsf{SI}$, $\forall n \in\left(\mathbb{I}_{K-1} \backslash \mathbb{E}_{P, K-1}\right)$.


\setlength{\textfloatsep}{10pt}
\begin{algorithm}[t]
\caption{Modulo-domain Sparse Channel Estimation}
\label{alg:FPrecovery}
\textbf{Input:}  $\{r_{\mathsf{SI}}[k], r_{\mathsf{SI},\lambda}[k] \}_{k=0}^{K-1}, T_p$, and $P$. \\
\textbf{Output:}  $\mathbf{h}_{\mathsf{SI}}$.
\begin{algorithmic}
\State 1) Compute $\widehat{\underline{\mathbf{r}}}_{\mathsf{SI}, \lambda} = \mathbf{V} \underline{\mathbf{r}}_{\mathsf{SI}, \lambda}$ (DFT).
\State 2) Define $[\mathbf{z}]_{n}=-[\widehat{\mathbf{r}}_{\mathsf{SI},\lambda}]_{n}, n \in \mathbb{I}_{K-1} \backslash \mathbb{E}_{P, K-1}$.
\State 3) Estimate number of folds, $M$, by forming a Toeplitz or Hankel matrix from $[\mathbf{z}]_{n}$ \cite{Bhandari:2021:J} and then thresholding using \textit{second order statistic} of eigenvalues \cite{he2010detecting}.
\State 4) Estimate folds using \texttt{Prony}$(\mathbf{z}) \mapsto\left(\left\{\mu_m, \nu_m\right\}_{m=0}^{M-1}, \mathbf{f}\right)$.
\State 5) Construct $\widehat{\underline{\boldsymbol{\varepsilon}}}_r$ for $n \in [0, K-2]$.
\State 6) Estimate $[\widehat{\underline{\mathbf{r}}}_\mathsf{SI}]_n = [\widehat{\underline{\mathbf{r}}}_{\mathsf{SI}, \lambda}]_n + [\widehat{\underline{\boldsymbol{\varepsilon}}}_r]_n, n \in \mathbb{E}_{P, K-1}$.
\State 7) Calculate $\underline{\widehat{\boldsymbol{\gamma}}}$ with given reference signal $\gamma$. Compute element-wise division that is $\mathbf{d} ={\widehat{{\underline{\mathbf{r}}}}_\mathsf{SI}}/{\widehat{{\underline{\boldsymbol{\gamma}}}}}$, where $[\mathbf{d}]_{n} = e^{- \jmath \omega_0 \tau} [\mathbf{d}]_{n-1}$. 
\State 8) Calculate $q \triangleq {([\mathbf{d}]_{n} - [\mathbf{d}]_{n+1})} / {([\mathbf{d}]_{n-1} - [\mathbf{d}]_{n})}$.
\State 9) Compute $\widetilde{\tau} = - K \arctan(q) / (2 \pi)$, $\widetilde{A} = \mathscr{E}(|{\mathbf{d}} |)$, and $\mathbf{h}_\mathsf{SI} = \widetilde{A} \delta(k - \widetilde{\tau})$.
\State 10) Reconstruct $r_\mathsf{SI} = h_\mathsf{SI} \ast \gamma$ and subsequently subtract it from the received signal (r)
\end{algorithmic}
\end{algorithm}

With $\widehat{\boldsymbol{\varepsilon}}_\mathsf{SI}$ known, we can remove its contribution and obtain $ \widehat{{\underline{\mathbf{r}}}}_\mathsf{SI} = \widehat{\underline{\mathbf{r}}}_{\mathsf{SI}, \lambda} + \widehat{{\underline{\boldsymbol{\varepsilon}}}}_\mathsf{SI}$. Instead of recovering the signal ${\mathbf{r}}_\mathsf{SI}$, we follow a direct Fourier domain \uline{deconvolution} approach to estimate ${\mathbf{h}}_\mathsf{SI}$. To this end, note that,
\vspace{-3pt}
\begin{align}
\forall n \in\left(\mathbb{E}_{P, K-1}\right)+(K-1) \mathbb{Z}, \,\,
\widehat{{\underline{{r}}}}_\mathsf{SI} [n] =
\widehat{{\underline{{\gamma}}}} [n] A e^{-\jmath \omega_0 n \tau}
\vspace{-8pt}
\end{align}
The process of deconvolution now entails point-wise division,
\vspace{-3pt}
\begin{equation}
\label{eq:dp}
    d[n] \triangleq 
    \frac{
    {\widehat{{\underline{r}}}_\mathsf{SI} [n] }}
    {{\widehat{{\underline{\gamma}}}[n]}} 
    = A e^{-\jmath n \omega_0 \tau} \equiv 
    \widehat{h}_\mathsf{SI}[n].
    \vspace{-3pt}
\end{equation}
With $[\mathbf{d}]_n = d[n]$, we can now estimate $A = | {\mathbf{d}}|$ and more generally, via $\mathscr{E}(|{\mathbf{d}} |)$, where $\mathscr{E}$ denotes expectation operator. The unknown delay is estimated via, 
\vspace{-3pt}
\begin{equation}
\label{eq:10}
     [\mathbf{d}]_{n} / [\mathbf{d}]_{n-1} = 
     e^{- \jmath \omega_0 \tau}  \Rightarrow \tau = 
      \tfrac{1}{-\jmath \omega_0}\angle \left([\mathbf{d}]_{n} / [\mathbf{d}]_{n-1}\right).
\vspace{-3pt}
\end{equation}

\begin{figure}[t]
    \centering
    \includegraphics[width= 0.9\linewidth]{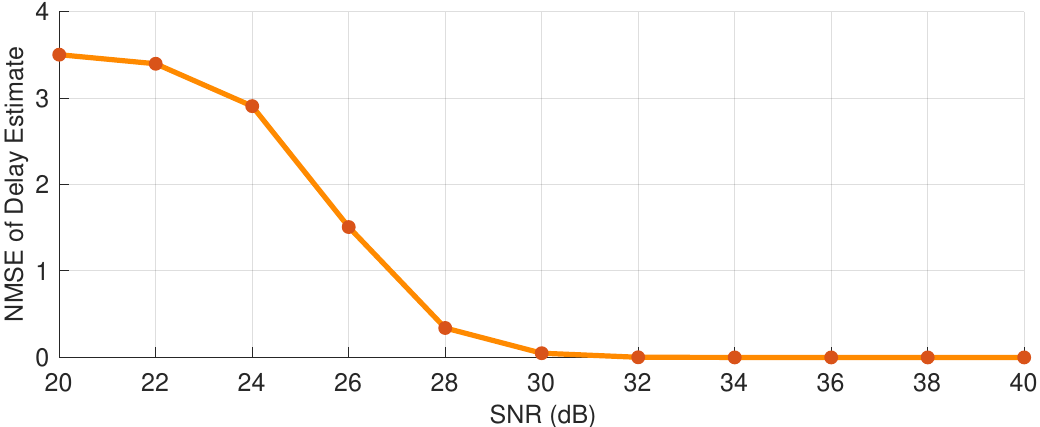}
    \caption{Normalized mean square error (NMSE) performance versus SNR with $20$ dB SI power.}
    \label{fig:NMSE}
    \vspace{-10pt}
\end{figure}

Furthermore, one may leverage measurement redundancy to obtain a robust estimate via least-squares solution. The procedural steps are outlined in Algorithm~\ref{alg:FPrecovery}. 
The computational complexity is approximately 
$\mathcal{O}(N_s \log(N_s) +  N_\textsf{Hankel}^3 +M^3)$, where $N_s$ is the input signal length, $N_\textsf{Hankel}$ is the Hankel matrix size, and $M$ is the number of exponentials to estimate. In comparison, adaptive filter-based algorithm (\eg  the normalized least mean square (NLMS) algorithm in \cite{ordonez2021full}) has the complexity $\mathcal{O}(N_{\mathsf{order}} I)$, where $N_{\mathsf{order}}$ is the filter filter order, $I$ is the iteration number. Although the proposed method has higher complexity, it effectively handles saturation and accurately reconstructs the self-interference (SI), as demonstrated in Section \ref{sec:result}. 
The performance of this algorithm measured by normalized mean square error (NMSE) is given in \fig{fig:NMSE}, which shows its noise robustness.
Having estimated $h_\mathsf{SI}$ via the proposed modulo-domain channel estimation method based on Theorem 1, we can reconstruct $r_\mathsf{SI}$ and subsequently subtract it from the received signal ($r$), to obtain the SoI, $r_\mathsf{SoI}$. 

\begin{theorem}[Modulo-domain Channel Estimation]
Let $g = h_\mathsf{SI} \ast \gamma$ where $h_\mathsf{SI}(t) = A\delta(t-t_\tau)$ is the unknown $1$-sparse channel, and $\gamma \in \mathcal{B}_\Omega$ is a known, $T_p$-periodic, transmitted signal. Given $K$ modulo samples $g_\lambda[k] = \MO{g(kT)}, T > 0$ folded at most $M$ times, the sufficient condition for estimation of $h_\mathsf{SI}$ from $\{g_\lambda[k]\}^{K-1}_{k=0}$ is that $T \leq \frac{T_p}{K}$ and $K \geq 2(M+2)$.
\end{theorem} 

\textit{Proof:} For the reconstruction Algorithm~2 to succeed, the exponential fitting part (\ie Prony's method) in step 2 requires 
\vspace{-8pt}
$$K \geq 2(P+M+1)
\vspace{-3pt}
$$ 
while step 8 (see (9)) requires that $P \geq 1$. With the period $\tau = KT$, the recovery condition thus is, $T \leq \frac{\tau}{2M+4}$ with $P = \lceil \frac{\Omega}{\omega_0} \rceil \geq 1$.



\section{Analysis of Quantization Noise}
\label{sec:analysis}
\begin{figure}[t]
\centering
		\includegraphics[width = 0.45\textwidth]{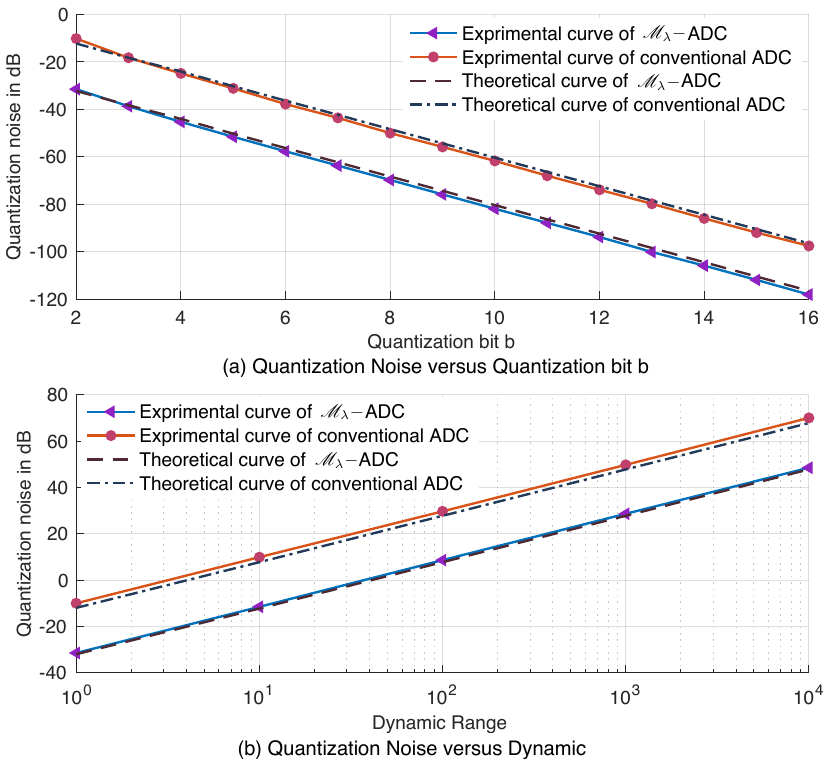}
		\caption{Quantization noise of $\mathscr{M}_\lambda$ and conventional ADCs versus (a) quantization $b$ when the DR and $\zeta = 0.1$ are fixed; (b) DR when the quantization $b=2$ and $\zeta = 0.1$ are fixed.}
		\label{fig:quant}
\vspace{-10pt}
\end{figure}
To understand the trade-off between bit budget and DR of both the \madc and conventional ADC, we analyze the quantization noise. Our theoretical analysis and simulations show that the \madc can achieve a better trade-off between bit budget and DR; the quantization noise of \madc is smaller compared to that of the conventional ADC. Since we focus on the quantization noise, we assume that DR, $\rho_\lambda$, matches the peak-to-peak amplitude of the signal $r$. Thus, there is no clipping error. 

The amplitudes of the single-carrier signal are subject to a uniform distribution, $r[k] \sim \mathcal{U}(0,1)$, thus the variance of the sample $r[k]$ is $\sigma_{r}^{2} = \rho_\lambda^2/12$. The variance of the quantization noise $\sigma_\mathsf{q}^{2}$ is given by
\vspace{-10pt}
\begin{equation}
    \sigma_\mathsf{q}^{2}=\frac{q_0^{2}}{12}=\sigma_{r}^{2} \ 2^{-2 b},
    \label{eq:q_c}
\vspace{-3pt}
\end{equation}
where $\qo = 2^{-b} \rho_\lambda$ is the step size of the mid-rise quantizer. Since the modulo samples follow a uniform distribution,  $r_\lambda[k] \sim \mathcal{U}(0, \zeta^2)$, the quantization noise of \madc is 
\vspace{-3pt}
\begin{equation}
    \sigma_{\mathsf{q, \lambda}}^2=\zeta^2\sigma_\mathsf{q}^2 = \zeta^2 \sigma_{r}^{2} \ 2^{-2 b},
    \label{eq:q_m}
\vspace{-3pt}
\end{equation}
where $\zeta = {\lambda} / {\| r \|_\infty}$ is the ratio between the modulo folding threshold and maximum value of the input signal. Hence, there is in theory a $20 \log _{10} (1/\zeta)$ dB reduction in quantization noise. 
Note that since the $\zeta$ should satisfy the constraint \cite{Bhandari:2020:Ja, liu2023lambda}
\vspace{-5pt}
\begin{equation}
    {\left( {T\Omega e} \right)^L} < \frac{\lambda }{{{{\left\| r \right\|}_{{L_\infty }}}}} = \zeta,
\vspace{-5pt}
\end{equation}
$\zeta$ cannot be $0$, and the gain in SQNR is bounded by 
\vspace{-5pt}
\begin{equation}
    \mathsf{SQNR}_\mathsf{gain} < 20 \log _{10}  { (\pi \Omega e)^{-L} }.
\vspace{-5pt}
\end{equation}

Based on \eqref{eq:q_c} and \eqref{eq:q_m}, we observe the quantization noise versus quantization bit $b$. Specifically, we vary the bit $b$ and set $\zeta = 0.1$ while ensuring that the dynamic ranges (DRs) of the \madc and conventional ADC are the same. As shown in \fig{fig:quant} (a), a $20$ dB reduction in quantization noise is observed. This is also consistent with the theory, as indicated by the dashed curves in the figures. From \eqref{eq:q_c} and \eqref{eq:q_m}, we can derive that the equivalent bit of \madc is, $b_\lambda = b + \log_2 \zeta$. This means that if we adopt $b_\lambda = 3$ quantization and set $\zeta = 0.1$, the quantization noise of \madc is approximately equivalent to that of a conventional ADC with $b=6$. Additionally, we observe the quantization noise versus DR with a fixed $b=2$ and $\zeta = 0.1$. Specifically, DR$=10$ means that DR is set to $10$ times the original DR. It is observed in \fig{fig:quant} (b) that the quantization noise of \madc is smaller (\eg $20$dB reduction when $\zeta = 0.1$) compared to a conventional ADC. Thus, with the same bit budget and DR, \madc has smaller quantization noise, facilitating a better trade-off between bit budget and DR compared to the conventional ADC.

\section{Numerical evaluation}
\label{sec:result}
This section focuses on the unique advantages of adopting the USF for full-stack FD architecture. According to the simulations, we show the saturation-free recovered (received) signal, reconstructed SoI and detection performance. 
We consider single-carrier quadrature phase shift keying (QPSK) based transmission, and the SI channel is modelled as a single path channel with delay. Additionally, the SI power is measured by $\mathsf{SIR} = 10 \log_{10} (P_\mathsf{SoI}/P_\mathsf{SI})$ dB. {We assume that no uplink data is transmitted during the SI channel estimation period, and perfect CSI of the uplink channel is known at the receiver. }
With respect to the recovery algorithm,
we follow $f_s \geq 2 \pi e f_\mathsf{Nyquist}$ to  reconstruct the signal in the simulation. This oversampling enables a fully digital-domain SI cancellation and a lower hardware complexity due to no AGC and VGA. We remind the reader that this is a loose bound, a near Nyquist-rate recovery is also possible \cite{bhandari2019identifiability}, specially with some side-information \cite{ordentlich2018modulo}, which is not assumed in this paper. 

\bpara{Recovery of received signal and SoI after SIC.}
\begin{figure}[t]
\centering
		\includegraphics[width = 0.45\textwidth]{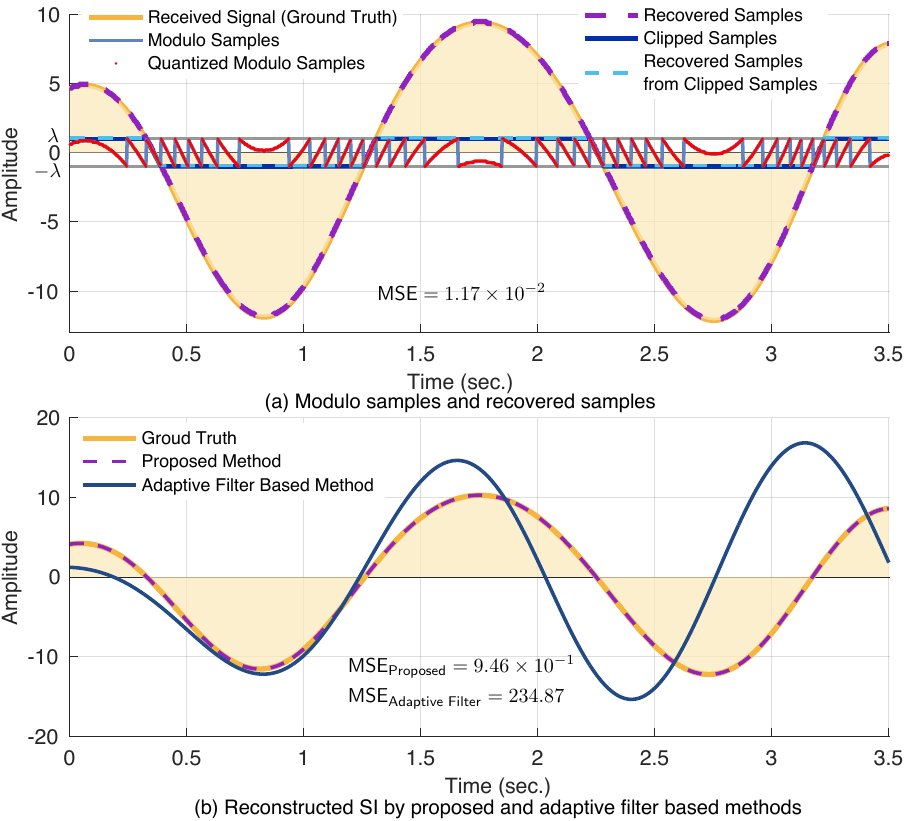}
		\caption{Simulation results with $4$-bit quantization, 1/SIR = $20$ dB, and SNR = $40$ dB of (a) modulo samples and recovery; (b) SI reconstruction by proposed SIC and adaptive filter based approach.}
		\label{fig:rxandsoi}
\vspace{-10pt}
\end{figure}
Since we consider the FD system, it is natural to work with signals such that $\| r \|_\infty \triangleq \max |r\rob{t}| \gg \lambda$. To prevent saturation, the received signal is folded by the \madcs. Subsequently, modulo-domain channel estimation as in Algorithm \ref{alg:FPrecovery} is adopted to estimate SI channel, $\mathbf{h}_\mathsf{SI}$. Then, the unfolded received signal $\widetilde{\mathbf{r}}$ is recovered from the modulo samples $\mathbf{r}_\lambda$. With the estimated SI channel and baseband transmit signal $\mathbf{\gamma}$ (reference signal), the digital-domain SIC is applied. Specifically, we reconstruct the SI signal $\mathbf{r}_\mathsf{SI}$, and subtract it from the reconstruction of the received signal $\widetilde{\mathbf{r}}$ to obtain $\mathbf{r}_\mathsf{SoI}$. Finally, the QPSK signal detection is executed on the $\mathbf{r}_\mathsf{SoI}$.

Considering low-resolution quantization (\ie $b=4$) with 1/SIR = $20$ dB (\eg $\lambda = 1,\| r \|_\infty = 10$), and SNR = $40$ dB, the reconstruction error of the received signal is $\mathsf{MSE} = 1.17 \times 10^{-2}$. As shown in \fig{fig:rxandsoi} (a), since the receiver ADC is saturated, the received signal in previous digital SIC method has clipping errors, and the information loss is irreversible. This will cause the failure of reconstruction as the curve ``Recovered Samples from Clipped Samples''. Thus, it results in failure of the subsequent process.
Additionally, as shown in \fig{fig:rxandsoi} (b), the adaptive filter-based NLMS approach under full DR (\ie no saturation) in \cite{ordonez2021full} cannot accurately reconstruct the SI signal even with no saturation. This is because the SI signal is large and fast-changing. As a result, the subsequent SoI recovery and signal detection fail. In contrast, the proposed method can accurately reconstruct the SI with an MSE of $9.46 \times 10^{-1}$. Consequently, the reconstruction error of SoI is $\mathsf{MSE} = 3.47 \times 10^{-2}$, and bit error rate (BER) is $7.47 \times 10^{-2}$. This comparison clearly demonstrates the advantages of our proposed method in terms of digital SIC performance, SoI reconstruction, and detection.

If we consider a larger SI power that is 1/SIR = $40$ dB, the reconstruction error of the received signal is $\mathsf{MSE} = 1.22 \times 10^{-2}$. The reconstruction error of the SoI is $\mathsf{MSE} = 7.039 \times 10^{-1}$. The BER is $1.572 \times 10^{-1}$. {Note that this marks the initial step in fully digital-domain SI cancellation, pioneering a new approach to addressing SI. Although both SI and system noise in practice can be more severe, future advancements in recovery methods will enable us to address them effectively.} We attribute the reconstruction with few bits to the fact that for a given bit budget, the \madc has a smaller quantization noise, where a $20 \log _{10} (1/\zeta)$ dB reduction is shown in Section \ref{sec:analysis}.

\section{Conclusion}

In this work, we presented a method for full-stack FD receiver architecture, utilizing the USF. We demonstrated the advantages of substituting the traditional ADC with a \madc. This replacement effectively avoids ADC saturation, enabling the capture of HDR signals. Additionally, the low quantization noise feature of the \madc allows for the use of few-bits quantization. Additionally, we introduced a novel approach for sparse channel estimation in the modulo domain, accompanied by a digital SI cancellation method. Our numerical results demonstrate that the proposed full-stack FD receiver structure can tackle with very strong SI {(\eg $40$ dB)} with few-bits quantization, and outperforms the previous approach in terms of digital SIC, SoI reconstruction, and detection performance. To summarize, our work enables: 1) fully digital-domain SI cancellation in FD system; 2) modulo-domain sparse channel estimation; 3) a low complexity implementation of FD system due to no VGA, AGC and analog-domain SIC circuitry.

\bibliographystyle{IEEEtran_url}
\bibliography{IEEEabrv,ref_new}

\end{document}